\documentclass[twocolumn,twoside,slac]{revtex4}
\usepackage{graphicx}
\usepackage{fancyhdr}
\begin{document}

\title{Atlas Data-Challenge 1 on NorduGrid}

%
\author{P.\ Eerola, B.\ K\'onya, O.\ Smirnova}
\affiliation{Particle Physics, Institute of Physics, Lund University, Box 118, 22100 Lund, Sweden}
\author{T.\ Ekel\"of, M.\ Ellert}
\affiliation{Department of Radiation Sciences, Uppsala University, Box 535, 75121 Uppsala, Sweden}
\author{J.\ R.\ Hansen, J.\ L.\ Nielsen, A.\ W\"a\"an\"anen}
\affiliation{Niels Bohr Institutet for Astronomi, Fysik og Geofysik, Blegdamsvej 17, Dk-2100 Copenhagen \O, Denmark}
\author{S.\ Hellman}
\affiliation{Department of Physics, Stockholm University, SCFAB, SE-106 91 Stockholm, Sweden}
\author{A.\ Konstantinov}
\affiliation{University of Oslo, Department of Physics, P.\ O.\ Box 1048, Blindern, 0316 Oslo, Norway and Vilnius University, Institute of Material Science and Applied Research, Saul\.etekio al.\ 9, Vilnius 2040, Lithuania}
\author{T.\ Myklebust, F.\ Ould-Saada}
\affiliation{University of Oslo, Department of Physics, P.\ O.\ Box 1048, Blindern, 0316 Oslo, Norway}

\begin{abstract}
The first LHC application ever to be executed in a computational Grid
environment is the so-called ATLAS Data-Challenge 1, more specifically, the 
part assigned to the Scandinavian members of the ATLAS
Collaboration. Taking advantage of the NorduGrid testbed and tools,
physicists from Denmark, Norway and Sweden were able to participate in
the overall exercise starting in July 2002 and continuing through the
rest of 2002 and the first part of 2003 using solely the NorduGrid
environment. This allowed to distribute input data over a wide area,
and rely on the NorduGrid resource discovery mechanism to find an
optimal cluster for job submission. During the whole Data-Challenge 1, more
than 2 TB of input data was processed and more than 2.5 TB of
output data was produced by more than 4750 Grid jobs.

\end{abstract}

\maketitle

\thispagestyle{fancy}

\section{Introduction}
In order to prepare for data-taking at the LHC starting 2007, ATLAS has
planned a series of computing challenges of increasing size and complexity
 \cite{dc1ref}.
The goals include a test of its computing model and the complete software 
suite and to integrate grid-middleware as quickly as possible. 
The first of these Data-Challenges, the Atlas Data-Challenge 1, has run in 
several stages in the second half of 2002 and in the first half of 2003. For
the first time, a massive worldwide production of many different physics
samples was run in a total of more than 50 institutes around the world.
NorduGrid was the Scandinavian contribution to this Data-Challenge.

In the following, we will describe the NorduGrid involvement in this
Atlas Data-Challenge 1. We start out by giving a description of the 
NorduGrid testbed and then move on to describe its tools seen from a 
user's perspective with some emphasis on how Atlas jobs are run
on NorduGrid. Then we describe the different stages in the Atlas
Data-Challenge 1 and the achievements of NorduGrid.
A detailed description of the NorduGrid architecture will be given in a
separate article~\cite{otherCHEP}.

\section{The NorduGrid testbed}
The aim of the NorduGrid project has from the start been to build and operate 
a production Grid in Scandinavia and Finland. The project was started in May 
2001 and has since May 2002 been running a testbed based on the architecture
discussed in \cite{archi}. Since July 2002, it has been used for production
by different groups as for example described in this article. The NorduGrid 
software is a light-weight and portable grid-middleware
that requires minimal changes to already existing non-grid resources
from the corresponding system administrators. This has allowed NorduGrid
to get adopted by many existing supercomputer centres
in the Nordic countries. The NorduGrid software runs on clusters with 
such different Linux distributions as Redhat 6.2, Redhat 7.2, Mandrake 8.0, 
Debian 3.0 (Woody) and others.

\begin{figure*}[t]
\centering
\includegraphics[width=100mm]{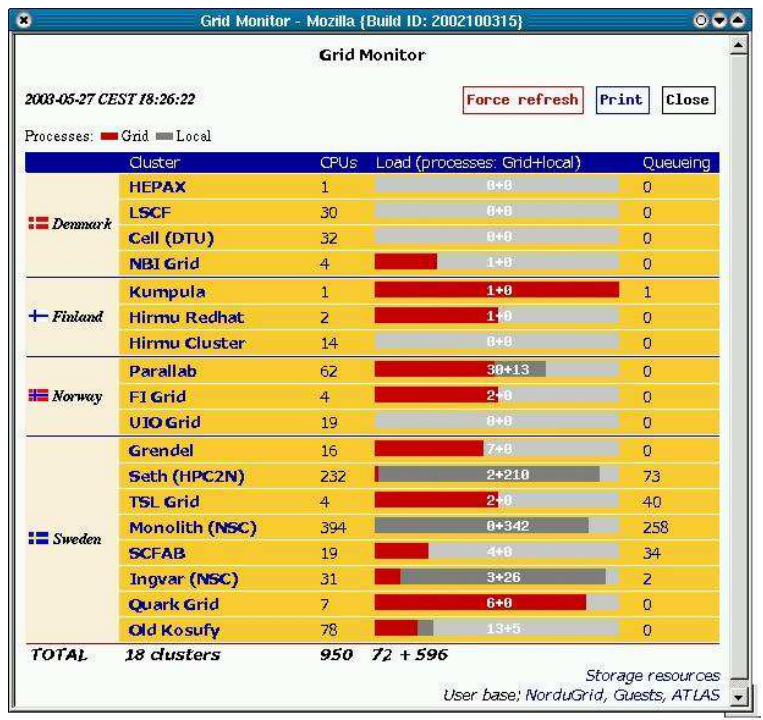}
\caption{A typical screenshot of the GridMonitor during production. The
number of CPU's for each cluster is given to the right of each cluster name. 
The number of running grid-jobs for each cluster is given by the red 
bar and the left number in the middle of the bar. The number of locally 
submitted (non-grid) jobs is given by the grey bar and the 
right number in the bar. The total number of clusters, CPU's, running
grid-job and locally submitted job are given at the bottom.}
\label{gridmon}
\end{figure*}

By now, the NorduGrid resources range from the 
original small test-clusters at the different physics-institutions to some of 
the biggest supercomputer clusters in
Scandinavia. It is one of the largest operational Grids in the world with
approximately 1000 CPU's available 24 hours a day, 7 days a week. The list
of clusters currently available on NorduGrid can at anytime be seen
on the NorduGrid GridMonitor (http://www.nordugrid.org/monitor/loadmon.php).
See also figure~\ref{gridmon}.

\section{The NorduGrid User Interface}
The NorduGrid User Interface consists of a set of command-line tools
used for submitting jobs, for querying the status of jobs, for fetching the 
output of jobs (if any), for killing and deleting jobs, for moving and copying
files between Storage Elements and Replica Catalogs and so on.
The complete list of commands can be seen in table~\ref{ng-commands} and
detailed information about these can be found in the NorduGrid User Interface 
manual~\cite{NGUI}.

To submit a job, the user has to specify his job-requirements in
the so-called eXtended Resource Specification Language,
xRSL~\cite{xRSL}.
xRSL is an extension of the Globus job-description language 
RSL~\cite{globusRSL} and is used to describe in a set of attributes 
all the information needed to run the job --- name of executable, arguments, 
input-files and so on. In addition, the user can specify a set of 
requirements that a given cluster
must satisfy to be able to run the job. These requirements include e.g.\ 
that a certain amount of disk space is available for the job on the cluster. 

One of the most important requirements is that specified by the
{\tt runTimeEnvironment}-attribute. This attribute provides support
for general software packages runtime environment configuration.
When a site-administrator installs a given software package
on his cluster, he can advertise this in the Information
System by creating a bash-script with the name of the package
in the so-called {\sl runtimedir}
on his cluster containing the necessary environment setup for the
software package. This script is then run before job-execution for each
job that requests this runtime-environment with the
{\tt runTimeEnvironment}-attribute. A simple example of a runtime 
environment configuration script for the Atlas-rpms provided by NorduGrid is 
shown later in figure~\ref{atlasenv}.

The user submits his grid-job to NorduGrid using the command ngsub.
During job-submission, the User Interface queries the NorduGrid Information
System for available resources and clusters and then queries each cluster 
to perform the check of the requirements specified by the user. If the xRSL
specifies that certain input-files should be downloaded from a Storage
Element or Replica Catalog, the User Interface contacts these to obtain
information about the input-files. Finally the User Interface performs the 
brokering according the information gathered and submits the job to the
chosen cluster.

From here on, the NorduGrid Grid Manager automatically takes care of the job 
--- handling the pre-processing (e.g.\ downloading input-files and so on), 
job-submission to the local PBS-system and post-processing (e.g.\ 
automatically uploading output-files to Storage Elements). The user can at 
anytime check the status of his submitted job using the command 
ngstat. A job can be in several different states during job-processing.
A list of possible states with a short explanation is shown in 
table \ref{states}.

When a job has finished, the user can retrieve those output-files that have 
not already been uploaded automatically to a Storage Element using the 
command ngget. Furthermore the user can at anytime get the standard output
of the job displayed using ngcat. This can be useful for checking whether
a job is running as intended. If it's not, the user can kill it using
ngkill and try again.

\begin{table*}[htb]
\centering
\begin{tabular}{|c|c|}
\hline
command & action \\
\hline
\verb#ngsub#    & Job submission \\
\verb#ngstat#   & Show status of jobs\\
\verb#ngcat#    & Display standard output of running jobs\\
\verb#ngget#    & Retrieve output from finished jobs\\
\verb#ngkill#   & Kill running jobs\\
\verb#ngclean#  & Delete jobs from a cluster \\
\verb#ngsync#   & Update the User Interface's local information about running jobs \\
\verb#ngcopy#   & Copy files to, from and between Storage Elements and replica catalogues\\
\verb#ngremove# & Delete files from Storage Elements and Replica Catalogs \\
\hline
\end{tabular}
\caption{\label{ng-commands}The NorduGrid User Interface commands.}
\end{table*}

\begin{table*}[bth]
\centering
\begin{tabular}{|c|l|}
\hline
state & meaning \\
\hline
ACCEPTED    & The job has been submitted to a cluster.\\
            & --- it is waiting to be processed by the Grid Manager.\\
PREPARING   & The Grid Manager is preparing the job\\
            & --- e.g.\ downloading input-files specified by the job-description.\\
INLRMS: Q   & The job has been submitted to the local PBS-system and is queued.\\
INLRMS: R   & The job has been submitted to the local PBS-system and is running.\\
FINISHING   & The job has finished running in the local PBS-system.\\
            & --- the Grid Manager is doing the post-processing.\\
FINISHED & The job has finished. The output of the job can be retrieved.\\
\hline
\end{tabular}
\caption{\label{states}Possible states of a NorduGrid grid-job.}
\end{table*}

\section{Atlas-software}
The different releases of the Atlas software are installed at CERN in a public
afs-directory under the ATLAS tree. Sites with afs-connection can use these 
Atlas releases directly but this solution is far from ideal especially since
the releases has for a long period been very Redhat 6.1 specific.

For Data-Challenge 1, official Atlas-rpms for the different releases
were distributed. These rpms were used by many sites with satisfaction.
But primarily because of the diversity of different Linux-distributions on
NorduGrid and the requirements of the different site-administrators, NorduGrid
decided to produce its own set of Atlas-rpms that one was able to build from 
scratch starting from a generic vanilla Linux-distribution. Indeed the
NorduGrid Atlas-rpms
\begin{itemize}
\item are buildable from scratch on many different Linux
distributions from the provided SRPMS.
\item are trivially relocatable.
\item have an install-area that allows a very simple
Atlas-{\tt runTimeEnvironment} to be setup.
\item come with an automatic install-script.
\end{itemize}

With the efforts of the different site-administrators on NorduGrid, these 
Atlas-rpms have been built and installed on many of the Nordu\-Grid sites.
Lately these Atlas-rpms have also been used in Chicago in phase 3 of
Data-Challenge 1 in connection with the Chimera project 
(see http://grid.uchicago.edu/atlaschimera/). In figure~\ref{atlasenv}
a typical runtime-environment script for Atlas-release 6.0.3 is shown.
The simplicity of this script is due to the install-area that these
Atlas-rpms are equipped with.

\begin{figure}[htb]
\begin{verbatim}
#!/bin/sh

ATLAS_ROOT=/usr/local/atlas/6.0.3
ROOTSYS=/usr/local/root/3.04.02
G4INSTALL=/usr/local/geant4/4.5.0.ref02
G4SYSTEM=Linux-g++

PATH=$ATLAS_ROOT/bin:$ROOTSYS/bin:$PATH

LD_LIBRARY_PATH=$ATLAS_ROOT/lib:$ROOTSYS/lib:\
$G4INSTALL/lib/$G4SYSTEM:$LD_LIBRARY_PATH

export PATH LD_LIBRARY_PATH ATLAS_ROOT ROOTSYS\
G4INSTALL G4SYSTEM
\end{verbatim}
\caption{Example of runtimeenvironment script for Atlas release
6.0.3 for the rpms provided by NorduGrid. Local adaptions of this script 
is run locally on the clusters before each Atlas-job.}
\label{atlasenv}
\end{figure}

Additionally for each Atlas-release, validation procedures for the
clusters were constructed. These consisted of a set of standard validation 
grid-jobs to test the Atlas-software-installation at each cluster. The
validation jobs were run on each cluster, before the site-administrators
were allowed to advertise the corresponding Atlas-{\tt runTimeEnvironment} 
in the Information System. This assured that the production would only be
performed on validated clusters.

\begin{figure*}[hbt]
\centering
\begin{verbatim}
&
(executable="ds2000.sh")
(arguments="1145")
(stdout="dc1.002000.simul.01145.hlt.pythia_jet_17.log")
(join="yes")
(inputFiles=("ds2000.sh" "http://www.nordugrid.org/applications/dc1/2000/dc1.002000.simul.NG.sh"))
(outputFiles=
   ("atlas.01145.zebra"
    "rc://dc1.uio.no/2000/zebra/dc1.002000.simul.01145.hlt.pythia_jet_17.zebra")
   ("atlas.01145.his"
    "rc://dc1.uio.no/2000/his/dc1.002000.simul.01145.hlt.pythia_jet_17.his")
   ("dc1.002000.simul.01145.hlt.pythia_jet_17.log"
    "rc://dc1.uio.no/2000/log/dc1.002000.simul.01145.hlt.pythia_jet_17.log")
   ("dc1.002000.simul.01145.hlt.pythia_jet_17.AMI"
    "rc://dc1.uio.no/2000/ami/dc1.002000.simul.01145.hlt.pythia_jet_17.AMI")
   ("dc1.002000.simul.01145.hlt.pythia_jet_17.MAG"
    "rc://dc1.uio.no/2000/mag/dc1.002000.simul.01145.hlt.pythia_jet_17.MAG")
)
(jobName="dc1.002000.simul.01145.hlt.pythia_jet_17")
(runTimeEnvironment="DC1-ATLAS")
(replicacollection="ldap://grid.uio.no:389/lc=ATLAS,rc=Nordugrid,dc=nordugrid,dc=org")
(CPUTime=2000)
(Disk=1200)
\end{verbatim}
\caption{A complete submission xrsl-script for dataset 2000 (phase 1)}
\label{xrsl-script1}
\end{figure*}

\section{Data-Challenge 1}
The simulations performed in Data-Challenge 1 was based on requests
from the High-Level-Trigger community and different physics communities.
These requests were divided into several datasets and each dataset was 
assigned either fully or in parts to different institutes participating 
in the Data-Challenge. The first phase, the simulation phase, running
from July to August 2002 consisted of large-scale ATLAS detector
simulations using the ATLAS physics simulation program, atlsim, from
Atlas release 3.2.1. In the second phase, running from December 2002 to 
January 2003, the already simulated events had to be piled up with 
minimum-bias events. This was also done in atlsim, with Atlas release
4.0.1, using a new highly efficient event mixing
procedure. The participating institutes were assigned the same partitions
as they had already simulated in phase 1.
The third phase consisted of large-scale reconstruction of the produced
data from the first two phases --- using Atlas releases 6.0.2 and
6.0.3.

In the first and second phase, NorduGrid was assigned 300 partitions of 
the dataset 2000 --- corresponding to 10 percent of the full dataset --- 
and the total dataset 2003 consisting of 1000 partitions. Before and
during phase 3 additional partitions from dataset 2000 and dataset 2001
were assigned and moved to NorduGrid for reconstruction.
In the following, we will describe in some detail the different phases.

\subsection{Phase 1 -- Simulation}
As mentioned NorduGrid was assigned 300 partitions of the dataset 2000.
The corresponding input-files containing the generated events to be 
simulated in the Atlas physics simulation program were 15 files each of 
a size of about 1.7 GB. Before the simulations, these input-files
were distributed to the clusters participating in this first phase and a 
corresponding {\tt runTimeEnvironment} DC1-ATLAS defining the environment 
setup for Atlas release 3.2.1 together with the location of the
input-files was put in place for each cluster.

A typical production xRSL for phase 1 dataset 2000 is seen in
figure~\ref{xrsl-script1}. The {\tt executable}-attribute specifies the
executable. In this case, the script {\tt ds2000.sh},
after linking in some data-files,
calls atlsim, which performs the actual detector simulations. 
The script {\tt ds2000.sh} is downloaded from the
URL given under {\tt inputFiles} and take the input-partition (in this
case 1145) as the {\tt argument}. The xRSL requires as discussed above
the {\tt runTimeEnvironment DC1-ATLAS} to ensure that the job is only
submitted to those clusters which have the Atlas release 3.2.1 installed
and the input-files present.

The standard output of the job will go into the file specified by 
the {\tt stdout}-attribute and the {\tt join} attribute requests that 
the standard error be merged with the standard output. Furthermore the 
name of the job is specified using the {\tt jobName}-attribute. The xRSL
also requests a certain amount of CPU time and diskspace through
the {\tt CPUTime} and {\tt Disk} attributes. The {\tt CPUTime} attribute
ensures e.g.\ that the User Interface chooses the right PBS-queue to submit
the job to on the chosen cluster and the job therefore is not killed 
when a certain amount of CPU time has been spent.

Each job produces a set of output-files. These are specified under
{\tt outputFiles} and will at the end of the job be automatically uploaded
by the Grid Manager to the location specified under
{\tt outputFiles}. In this case, the output-files specified will be
uploaded to a physical location registered in the NorduGrid Replica
Catalog defined by the {\tt replicaCollection} attribute, so that on request 
from the Grid Manager, the Replica Catalog
resolves e.g.\ {\tt rc://dc1.uio.no/log} to
{\tt gsiftp://dc1.uio.no/dc1/2000/log} and uploads the file to that
Storage Element. Similarly with the other files.

\begin{figure*}
\centering
\includegraphics[width=100mm]{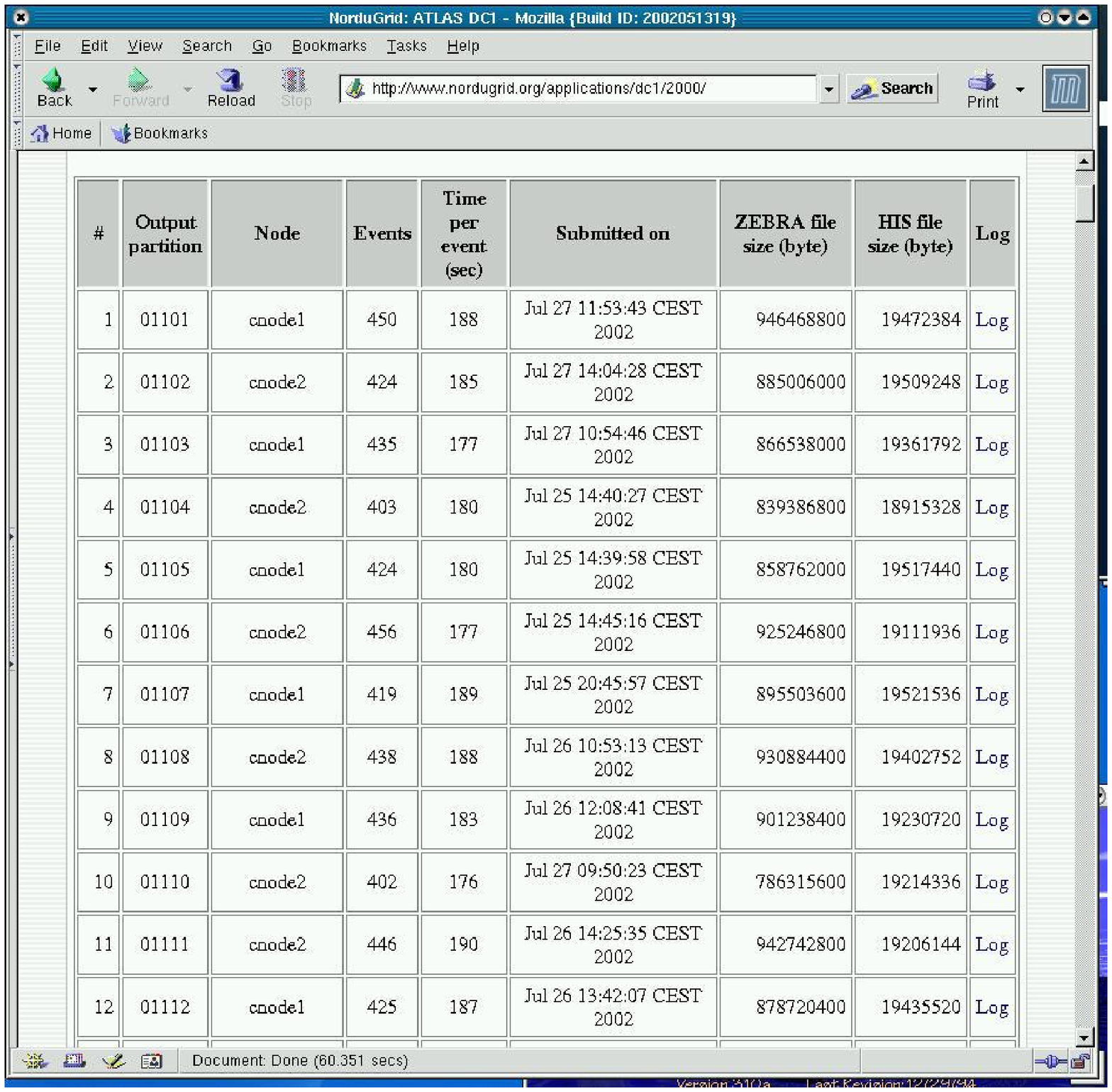}
\caption{A snapshot of the webpage querying the Replica Catalog for 
output-files of dataset 2000 during phase 1. For each partition, the 
production node, the number of fully simulated events, the CPU time spent 
per event, the submission time, the output sizes of the zebra- and 
histogram-files and the logfile are be shown. A red exclamation mark
is shown for bad partitions.}
\label{web2000}
\end{figure*}

The 300 jobs for dataset 2000 in this first phase used around 220
CPU-days and produced $5\times 300$ output-files with a total size
of about 320 GB. All output-files were uploaded to a NorduGrid Storage Element
as discussed above. A web page querying the Replica Catalog for output-files
was setup to allow for an easy browsing and check of log-files,
output-sizes, error-codes and so on. This provided a very easy check
of whether something had gone wrong during the job-processing. A 
screen-shot of this is shown in figure~\ref{web2000}

\begin{figure*}
\centering
\includegraphics[width=100mm]{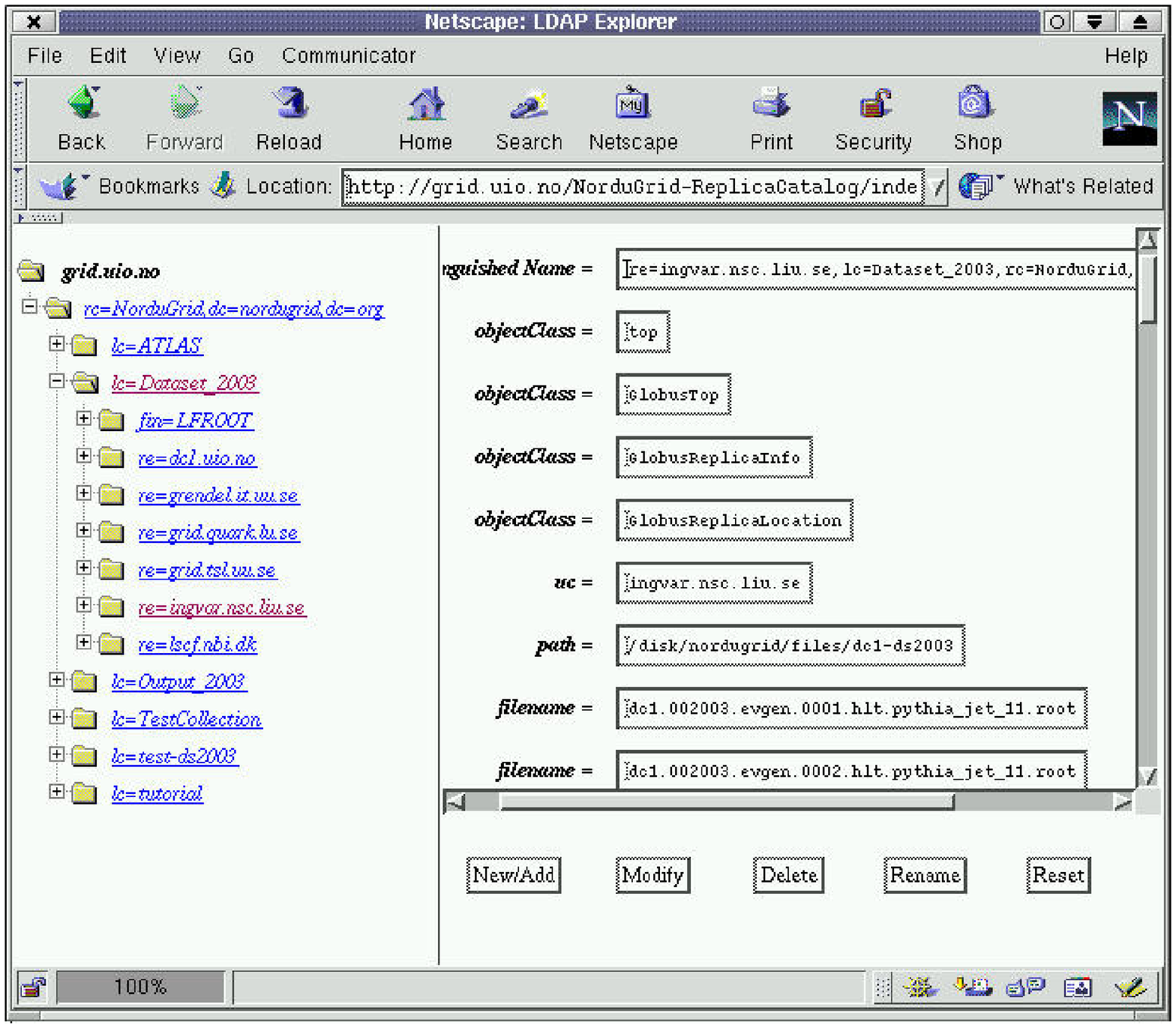}
\caption{A snapshot of the Replica Browser showing the dataset 2003
input-files as registered in the Replica Catalog.}  
\label{rc2003}
\end{figure*}

For dataset 2003, the strategy with the input-files was somewhat different. 
The input data to be processed consisted of 100 files with a total
volume of 158 GB. Not all sites in NorduGrid was able to accommodate such
a big amount of input-files, and it was therefore decided to distribute
only subsets of these. All the distributed input sets were then registered
into the Replica Catalog so that at the job-submission stage, the broker
could query the Replica Catalog for clusters having the necessary
input-file for that job. The job would then preferably be submitted to
such clusters. However, if all clusters with the required input-file
were full, the job would be submitted somewhere else, and the Grid
Manager would use the information in the Replica Catalog and proceed to
download the input-file into the local temporary cache on the
cluster. A snapshot of
the Replica Catalog with the dataset 2003 input-files registered is
shown in figure~\ref{rc2003}.

Otherwise the simulation were in principle completely similar to dataset
2000. For this part, there were 1000 jobs using about 300 CPU-days and
producing approximately 442 GB of output. All output-files were uploaded
to the designated NorduGrid Sto\-rage Element and registered into the
Replica Catalog. In total for phase 1, NorduGrid produced 780 GB of data 
in a total of 1300 jobs.

\subsection{Phase 2 -- Pileup}
 From a grid point of view, all 3 phases were very similar
--- in principle only the application changed. The biggest difference
from phase 1 to phase 2 was that the input-files of phase 2 were the
output-files of phase 1. This meant that the Grid Manager during 
job-preparation had to download each input-file from the designated
NorduGrid Storage Element. It turned out that this did not pose any
significant problems. The minimum-bias files were distributed as
in phase 1 to the different sites before the simulations. Again -- if a
chosen site did not have all the required minimum-bias files, the Grid Manager
would proceed to download these as well into the local cache.

In this phase 1300 jobs were processed having about 780 GB of input-files
and producing about 1080 GB of output-files. All output-files were again
automatically uploaded to a designated Storage Element and registered
into the Replica Catalog.

\subsection{Phase 3 -- Reconstruction}
During phase 3, NorduGrid was assigned additionally 400 partitions from dataset
2000 and 750 partitions from dataset 2001 to be reconstructed together
with the 300 + 1000 partitions from dataset 2000 and 2003 already present
on NorduGrid. The input-files for these were transferred directly
from Russia and the US to a
NorduGrid Storage Element using the ngcopy-program
and used in the simulations from there.

In the reconstruction phase, an external mysql noise database situated 
either at CERN or at BNL was used for
the first time to provide noise for the LArCalorimeter-system. This
initially posed a number of problems since most clusters in NorduGrid
do not allow for internet-connection from the worker nodes and a
reconstruction job would thus die because it could not contact the
external database. This was quickly solved though so that a job
could download a standalone version of the database instead and use that
in the reconstruction.

With this, no further problems were encountered and the
corresponding partitions have now all been reconstructed. In this phase
the NorduGrid share has been between 15 and 20 percent.

\section{Conclusion}
NorduGrid, as the Scandinavian contribution, has contributed 
substantially to the Atlas Data-Challenge 1 in all 3 phases. Important
lessons about the NorduGrid middleware has been learned during these
stress tests which has been used to extend the stability, flexibility
and functionality
of the software and NorduGrid itself. The software is now being adopted by
several computer centres throughout the Scandinavian countries which shows
exciting prospects for NorduGrid in Scandinavia in the future.

\begin{acknowledgments}
The NorduGrid project was funded by the Nordic Council of Ministers
through the Nordunet2 programme and by NOS-N. The authors would like to
express their gratitude to the system administrators across the Nordic
countries for their courage, patience and assistance in enabling the
NorduGrid environment. In particular, our thanks go to Ulf
Mj\"ornmark and Bj\"orn Lundberg of Lund University, Bj\"orn Nilsson of
NBI Copenhagen, Niclas Andersson and Leif Nixon of NSC Link\"oping,
\AA ke Sandgren of HPC2N Ume\aa\ and Jacko Koster of Parallab, Bergen.
\end{acknowledgments}


\end{document}